\documentclass[twocolumn,showpacs,preprintnumbers,prb,fleqn,floatfix]{revtex4}

\usepackage{graphicx}
\usepackage{dcolumn}
\usepackage{bm}
\begin{document}

\title{{Quasiparticle density of states in Bi$_2$Sr$_2$CaCu$_2$O$_{8+\delta}$
single crystals probed using tunneling spectroscopy at ultra-low
temperatures in high magnetic fields.}}
\author{S. I. Vedeneev$^{1,2}$ and D. K. Maude$^{1}$}
\affiliation{ $^1$ Grenoble High Magnetic Field Laboratory,
Max-Planck-Institut f\"{u}r Festk\"{o}rperforschung and Centre
National de la Recherche Scientifique,
B.P. 166, F-38042 Grenoble Cedex 9, France \\
$^2$ P.N. Lebedev Physical Institute, Russian Academy of Sciences,
119991 Moscow, Russia}

\date{\today }

\begin{abstract}
Break-junction tunneling spectroscopy at temperatures 30-50 mK in
high magnetic field is used to directly probe the quasiparticle
density of states within the energy gap in a single crystal Bi2212
high-$T_c$ superconductor. The measured tunneling conductances
$dI/dV(V)$ in the subgap region have a zero flat region with no
evidence for a linear increase of the density of states with
voltage. A number of tunnel break-junctions exhibited $dI/dV(V)$
curves with a second energy gap structure at the average magnitude
2$\Delta _{p-p}/e=13$ mV. Our data cannot be explained by either a
pure $s$ pairing or a pure $d_{x^2-y^2}$ pairing.
\end{abstract}

 \maketitle

\section{Introduction}

A careful investigation of the density of states (DOS) for
quasiparticle excitations in high-$T_c$ copper oxide
superconductors is essential for the understanding of the
mechanism responsible for superconductivity. Tunneling
spectroscopy has played an important role in verifying BCS theory
in conventional superconductors. The conductance of a tunnel
junction is directly proportional to the quasiparticle density of
states. The fine structure found in the tunneling conductance at
voltages above the energy gap is a direct proof that the
electron-phonon interaction is the coupling mechanism for
superconductivity in conventional superconductors. It is widely
accepted that the high-$T_c$ superconductors (HTS) have
$d_{x^2-y^2}$ symmetry of the order parameter. Among the HTS,
Bi$_2$Sr$_2$CaCu$_2$O$_{8+d}$ (Bi2212) has often been studied by
the tunneling method. Although the tunneling characteristics
obtained with Bi2212 single crystals show features related to the
superconducting energy gap, the shape of the tunneling conductance
$dI/dV$ vs $V$ is far from the ideal BCS density of states. The
conductance curves reveal a strong broadening of the
superconducting-gap structure, with a nonzero contribution at zero
voltage and a linear increase in the subgap region (cusp-like
feature) (see e.g. Ref.\onlinecite{Kit96,Hoog03}). In addition, an
unusual feature frequently observed in the tunneling data of
Bi2212 is the ``dip-hump'' structure beyond the gap edge which is
close to the resonant spin excitation energy.\cite{Zas01}

In superconducting tunnel junctions, the difference between $d$-
and $s$-wave symmetry leads to a significant change in the
current-voltage ($I(V)$) characteristics. For an intrinsic
unshunted $s$-wave superconductor-insulator-superconductor (SIS)
junction, the zero-bias conductance should be exponentially small
at low temperatures (especially in case of superconductors with
high $T_c$). In contrast, if the order parameter has nodes and
$d_{x^2-y^2}$ symmetry, the $I(V)$ characteristics of the $d$-wave
SIS junction in the vicinity of the zero bias should have a finite
slope and thus a nonzero conductance at zero voltage (see e.g.
Ref. [\onlinecite{Kit96}]). Experimental tunneling characteristics
of $d$-wave SIS junctions do not agree with the theoretical
predictions. Won and Maki \cite{Won94} found that at low
temperatures the subgap tunneling conductance $dI/dV$ should
increase almost quadratically with $V$. Whereas, Yamada and Suzuki
\cite{Yam02} have recently shown that for coherent tunneling, the
tunneling conductance of a $d$-wave SIS junction is distinctly
different when compared to incoherent tunneling. The subgap region
of the coherent SIS tunneling conductance is almost linear in $V$.
They find that the quasiparticle tunneling in a Bi2212 mesa is
mostly coherent. In general, the symmetry and pairing mechanism of
the superconducting state remains controversial. For this reason,
it is important to measure the subgap tunneling DOS of SIS Bi2212
junctions at very low temperatures and at high magnetic fields.
While the results of measurements of the superconducting energy
gap in SIS junctions are insensitive to thermal broadening,
measurement in the mK region are required to rule out thermal
excitations as the origin of both the excess conductance at zero
voltage and the broadening of the conductance peak.

Magnetic field is essential, firstly, in order to suppress the
Josephson current, which makes the measurement of the subgap
tunneling conductance difficult. Secondly, magnetic field is very
useful to distinguish relevant information in tunneling data from
anomalous features due to critical current effects in weak links
of the junction and the Bi2212 single crystal. Finally, it is
necessary to study the effect of the magnetic field on the subgap
and gap structures in the tunneling spectra.

Over the last few years the local DOS in Bi2212 has often been
studied spectroscopically using a scanning tunneling microscope
(STM). During the process of the mechanical cleaning, the crystal
breaks between the BiO planes of two adjacent half-unit
cells.\cite{Lindberg} In the case of tunneling measurements
performed on the BiO plane, the large separation between the tip
and the pair of CuO$_2$ planes (greater than 10 \AA) precludes the
possibility of tunneling directly into these planes. In this
situation, the electronic states of planes other than CuO$_2$ play
a role in the tunneling process and must be taken into
account.\cite{Misra}  In contrast to STM, in a break-junction
fabricated on Bi2212 single crystals by using a precision setup,
the tunneling can occur along CuO$_2$ planes because many plane
edges are created. It should be pointed out that since Bi2212
single crystals can be easily cleaved, the possibility exists that
the crystal shears along a plane rather than breaking cleanly,
leading to the formation of a $c$-axis junction, but we will argue
later that the tunneling in our break junctions is most likely to
be in the $ab$-plane. There are a number of articles that have
shown that a mechanically controllable break junction is one of
the better tunnel systems (see e.g. references
\onlinecite{Mandrus93,Post,Zas01}). Breaking the crystal at
cryogenic temperatures and in high vacuum (or in an inert
cryogenic fluid) guarantees two atomically clean surfaces, thereby
minimizing surface contamination. The distance between these
surfaces can be mechanically controlled. Increasing the distance
results in a smaller junction, which is eventually reduced to only
a few atoms. It is possible to form a tunnel junction with a
vacuum barrier between the two foremost atoms.\cite{Post}
Although, in view of the technical complexity, this method has not
been extensively applied, it has provided reliable results so far,
especially in the case of Bi2212 single crystals. At the present
time this is the only method which allows a tunneling
investigations of Bi2212 single crystals at ultra-low
temperatures. In this paper, we describe the first experimental
study of the tunneling DOS in copper oxide HTS at temperatures
30-50 mK in high magnetic fields up to 26 T using high-quality
break-junctions fabricated on Bi2212 single crystals. These
investigations are an extension of our previous tunneling studies
of the Bi-compound HTS.\cite{Ved94,Vedeneev01}

\section{Experiment}

The Bi2212 single crystals were grown by a KCl-solution-melt free
growth method.\cite{Gorina93,Vedeneev01} It is known, that
overdoping or underdoping of Bi2212 can be achieved by changes in
the oxygen content or by cation substitutions.\cite{Villard, Ooi}
The first method requires an annealing in oxygen or argon at very
high temperatures. A careful characterization of the annealed
samples reveals that changes in $T_c$ are always accompanied by a
severe degradation of the sample quality. Kinoda \textit{et al.}
\cite{Kinoda03} have showed also that the annealing in oxygen can
substantially increase the gap inhomogeneity. For this reason we
have used only high quality \textit{as grown} single crystals in
which, the substitution of trivalent Bi for divalent Sr during
growth reduces the hole concentration in the CuO$_2$ planes.\cite
{Harris97} As the Bi/Sr ratio increases, the number of holes doped
into the system decreases, which therefore pushes the system
towards the hole-underdoped regime. Very recently Eisaki
\textit{et al.} \cite{Eisaki} have shown that as the Bi/Sr ratio
in Bi$_{2+x}$Sr$_{2-x}$CaCu$_{2}$O$_{8+\delta}$ single crystals
approached 1, $T_c$ increases from 82.4 K for $x=0.2$ to 91.4 K
for $x=0.07$, 92.6 K for $x=0.04$, and eventually to 94.0 K.
Bi2212 single crystals with the excess Bi have repeatedly been
used in tunneling study of temperature and doping dependence of
the superconducting gap and pseudogap (see e.g. Ref.
\onlinecite{Matsuda}).

\begin{figure} %figure 1
\includegraphics[width=0.9\linewidth,angle=0,clip]{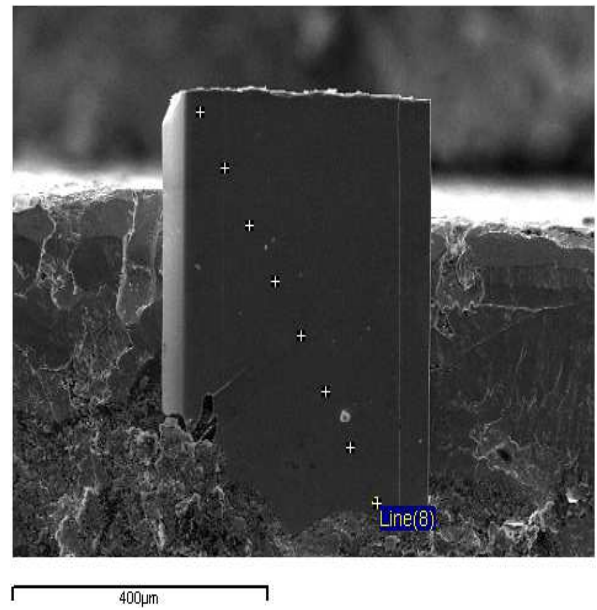}
\caption{\label{fig1} The scanning electron micrograph of a
crystal fragment where the composition measurement points are
denoted by crosses}.
\end{figure}

The quality of the crystals was verified by measurements of the dc
resistance, ac susceptibility, X-ray diffraction and scanning
electron microscopy. The single crystal showed X-ray rocking
curves with a width of about $0.1^{\circ}$ demonstrating the high
quality and high homogeneity of the samples. The values of the
residual resistivity in our crystals were comparable with the
lowest value previously reported for Bi2212. (see e.g. Ref.
\onlinecite{Watanabe97}). The composition of the crystals was
studied using a Philips CM-30 electron microscopy with a Link
analytical AN-95S energy dispersion X-ray spectrometer \cite{
Gorina05}. The actual cationic compositions of each crystal
investigated were measured at several different places on the
crystal and the scatter in the data was less than 2\%. In
Fig.~\ref{fig1} we show a scanning electron micrograph of a
crystal fragment where the composition measurement points are
denoted by crosses.  Previously, we have measured the Hall
coefficient in several crystals and found the nearly linear
relation between the excess Bi and the hole concentration $p$. Our
samples showed the well-known parabolic behavior for $T_c(p)$.

Since the deviation from conventional superconductivity should be
most pronounced in the underdoped regime, for the present study,
we have chosen three underdoped \textit{as grown} single crystals
Bi$_{2.22}$Sr$_{1.55}$Ca$_{1.17}$Cu$_{2.01}$O$_{8+\delta}$ with
$T_c$ = 84 K and transition width $\Delta T_c = 1.5$ K. The same
hole concentration for the investigated samples $p=0.125$ has been
obtained from the Bi/Sr ratio as well as from the empirical
relation $T_c/T_{c,max} = 1-82.6(p-0.16)^2$ which is satisfied for
a number of the HTS.\cite{Tal95} We used the fact that the
substitution of trivalent Bi for divalent Sr in the Bi-compounds
reduces the hole concentration in the CuO$_2$
planes.\cite{Harris97}

\begin{figure} %figure 2
\includegraphics[width=0.9\linewidth,angle=0,clip]{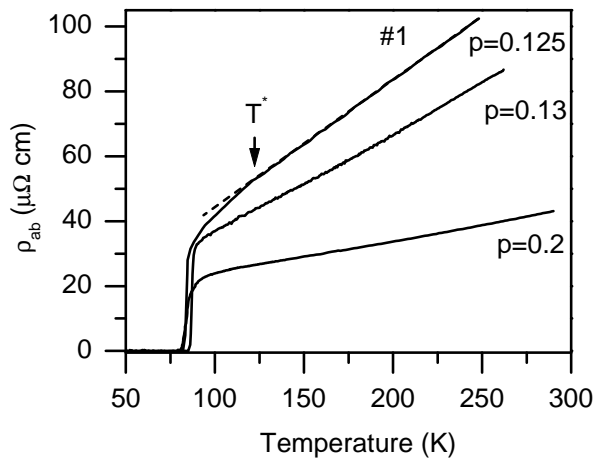}
\caption{\label{fig2} The temperature dependence of the in-plane
resistivity $\rho_{ab}$ for one of the investigating underdoped
crystals ($\#1$, $p=0.125$). For comparison, we show also
$\rho_{ab}(T)$ for an overdoped ($p=0.2$) single crystal with the
same $T_c = 84$ K (midpoint) as well as the data for a slightly
underdoped sample with $T_c = 87$ K ($p=0.13$)}.
\end{figure}

In order to demonstrate that the investigated samples are
underdoped, in Fig.~\ref{fig2} we show the temperature dependence
of the in-plane resistivity $\rho_{ab}$ for one of the
investigating crystals ($\#1$, $p=0.125$). For comparison, we show
also $\rho_{ab}(T)$ for the overdoped ($p=0.2$) single crystal
with the same $T_c = 84$ K (midpoint) as well as the data for the
slightly underdoped sample with $T_c = 87$ K ($p=0.13$). One can
see that as in all Bi-compound HTS, the magnitude of
$\rho_{ab}(T)$ increases with decreasing carrier concentration. A
typical $T$-linear behavior and a slightly upward curvature of
$\rho _{ab}$ are also seen in the slightly underdoped and the
overdoped samples. Whereas for the underdoped sample ($p=0.125$),
$\rho _{ab}$  deviates from high-temperature $T$-linear behavior
at a characteristic temperature $T^{*}$ (indicated by the arrow in
Fig. ~\ref{fig2}), as would be expected. This temperature in Ref.
[\onlinecite{Watanabe00}] was identified as the pseudogap closing
(opening) temperature $T^{*}$. This is a further proof that we are
measuring truly underdoped samples.

The dimensions of the investigated crystals were $\simeq 1~mm
\times (0.5-1)~mm \times (1-3)~\mu$m. A four-probe contact
configuration, with symmetrical positions of the low-resistance
contacts ($<1\Omega $) on both $ab$-surfaces of the sample was
used. For the tunneling measurements, the sample was cooled to 30
mK at zero magnetic field. The tunnel junction was then fabricated
\textit{in situ} using a mechanical break junction technique. With
the geometry of setup used, the break junctions forms so that
tunneling occurs in the $ab$-plane. The $I(V)$ characteristics and
differential conductances $dI/dV$ as function of $V$ were measured
using phase-sensitive detection techniques. We were able to
fabricate a large number of tunnel break-junctions at different
places along the initial break with resistances from 100 $\Omega$
to 120 k$\Omega$ (for bias voltages around $300$ mV) . Each
junction was stable with reproducible characteristics at different
magnetic fields. With our break-junction setup the magnetic field
was oriented parallel to the CuO$_2$ planes to an estimated
accuracy of $\approx 1^{\circ}$.\cite{Ved94}

In a previous investigation we have measured break junction
tunneling together with the $ab$-plane and out-of-plane
resistivities for the same Bi2212 crystal at different
temperatures (4.2 - 250 K) for magnetic fields up to 20 T oriented
parallel to the c-axis.\cite{Ved94} In a magnetic field
$\rho_{ab}(T)$ reveals a significant increase in the width of the
superconducting transition, while the value $T_c$ for the onset of
superconductivity remained constant. The data for $\rho_c(T)$
shows evidence for a strong suppression of the superconductivity
along the $c$-axis in a magnetic field of 20 T. The observed
decrease of the $T_c$ value by a factor 2 indicates a significant
reduction of the energy gap. In contrast, for all tunnel junctions
investigated the energy gap is almost independent of magnetic
field ($H\| c$-axis)$\leq 20$ T).\cite{Ved94} For this reason, we
can formally state that the observed gap spectrum for our break
junctions is related to the $ab$-plane superconducting energy gap
and therefore to tunneling in the $ab$-plane. At the same time, we
are unable to say anything about the tunneling direction in the
$ab$-plane itself. For example, it was shown in Ref.
[\onlinecite{Wei}] that the shape of spectra measured using point
contact tunneling spectroscopy of YBa$_2$Cu$_3$O$_7$ within the
$ab$-plane in the [100] and [110] directions were quite different
although the extracted magnitudes of the energy gap were in close
agreement. The exact knowledge of the tunneling direction is
essential for point contact tunneling spectroscopy where the main
contribution to the quasiparticle current is given by Andreev
reflection, which strongly depends on the wave vector. However,
this is not the case for tunnel junctions. Hoogenboom \textit{et
al.} \cite{ Hoog03} have shown on the basis of a detailed analysis
of a large number of tunneling and ARPES spectra that the
tunneling spectroscopy probes states along the entire Fermi
surface.

\section{Break junction tunneling spectroscopy}

\begin{figure} %figure 3
\includegraphics[width=0.7\linewidth,angle=0,clip]{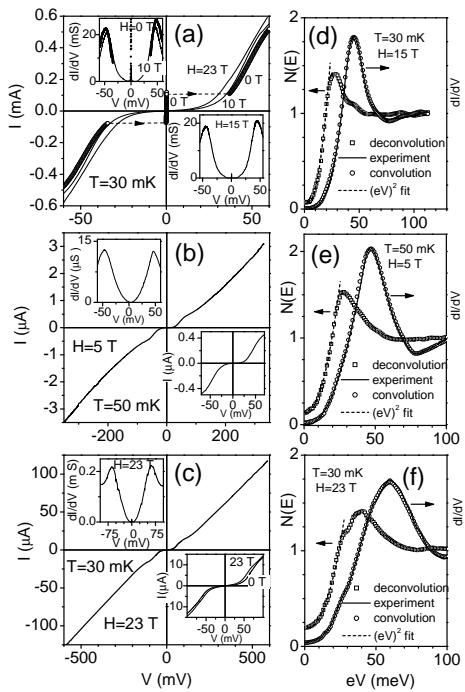}
\caption{\label{fig3} $I(V)$ characteristics for the (a) 100
$\Omega$-, (b) 115 k$\Omega$-, and (c) 6 k$\Omega$-resistance
tunnel junctions at $T=30$ and 50 mK, in different magnetic fields
(sample $\#1$). The insets in (a) and the upper insets in (b,c)
show the corresponding tunneling conductances data $dI/dV(V)$ in
the subgap region. The lower insets in (b,c) show the same $I(V)$
characteristics in the gap region. (d-f) the data points (squares)
represent the density of states $N(E)$ obtained from the
deconvolution of Eq.(\ref{eq1}) (see text)}.
\end{figure}

Fig.\ref{fig3}(a-c) shows representative $I(V)$ curves together
with the tunneling conductances data $dI/dV(V)$ for a 100
$\Omega$-, 115 k$\Omega$-, and a 6 k$\Omega$ tunnel junction,
respectively, at $T\simeq 40$ mK, for different magnetic fields
(sample $\#1$). All the $I(V)$ characteristics exhibit
characteristic features typical for superconducting tunnel
junctions, with a flat region around zero bias, consistent with
the expected zero tunneling conductance (the insets in
Fig.\ref{fig3} (a-c)), and a well-defined sharp increase in the
tunnel current near $\pm (45-65)$ mV connected with the
superconducting gap. At large bias voltages up to $\pm 0.6$ V, the
$I(V)$ characteristics are linear with small deviation (increasing
conductance) as expected for junctions with a good tunnel barrier
without current leakage.\cite{Wolf85} The $I(V)$ characteristics
for the low-resistance tunnel junction in the subgap region
(Fig.\ref{fig3}(a)) are also typical for low-resistance Josephson
junctions (the arrows indicate the direction of the bias current
sweep during the measurements). The $I(V)$ curves of this type
were hysteretic (versus bias sweep direction) and have the slope
of this switching. This is evident even from (Fig.\ref{fig3}(a)).
In accordance with the direction of the bias current sweep, the
Josephson current in zero magnetic field equals -76 and +120
$\mu$A. In our Bi2212 break-junctions we have observed the
expected Fraunhofer-like dependence of the Josephson critical
current as a function of  magnetic field. The large period of
Fraunhofer pattern points to the excessively small size of the
tunnel junctions. After the suppression of the Josephson current
in a magnetic field, the tunneling conductances $dI/dV(V)$ of
these junctions have a well-defined large zero flat region around
zero bias voltage (the lower inset in Fig.\ref{fig3}(a)). The high
resistance junctions do not show a zero-bias peak even at $H= 0$
T.

It should be pointed out that Joule heating effects could be
important at the ultra low temperatures, especially for low
impedance junctions. Previously we have studied Joule heating
effects in break-junctions fabricated on Bi2212 \cite{Ved94} and
Bi$_2$Sr$_2$CuO$_{6+\delta}$ (Bi2201) \cite{Ved01} single
crystals. It was found that the shape of the tunneling conductance
$dI/dV(V)$ and the value of the zero-bias $dI/dV$ are very
sensitive to a heating of the crystal region near the tunnel
barrier. The $dI/dV(V)$ characteristics in our data
(Fig.\ref{fig3}(a-c))  show no evidence whatsoever of heating. The
linearity of the current-voltage characteristics at high bias
voltages (up to 0.6 V) indicates that the heating caused by the
current injection in our measurements is negligible. The shape of
the tunneling conductance $dI/dV(V)$ is unchanged despite a
variation in the junction resistance of three orders of magnitude.

The measured tunneling conductances in the subgap region at
$V>|10|$ mV can be fitted quite well to curves which are quadratic
in $V$. The tunneling conductances in case of the high-resistance
tunnel junctions in Fig.\ref{fig3}(b,c) in the subgap region do
not show a large zero flat region. We have similar characteristics
to those shown in Fig.\ref{fig3}(a-c) for several tens of tunnel
break-junctions formed on each of the three single crystals.
\emph{We have never observed, either a linear increase of the
tunneling conductances with voltage in the subgap region, or a
nonzero zero-bias conductance for temperatures $\simeq 50$ mK.}
The conductance spectra with a flat bottom seems at first sight to
be consistent with $s$-wave rather than $d$-wave
superconductivity. It should be pointed that the nonlinear subgap
tunneling conductance in the vicinity of the zero bias has been
observed previously in Bi2212 \cite{Mandrus93,Ved94} and Hg-based
cuprates \cite{Chen94,Jeong94} at 4.2 K, but the magnitudes of the
background conductances at zero voltage were sizable, suggesting a
significant density of states of quasiparticles at zero energy.

Our results are nevertheless puzzling since the tunneling
conductances in the subgap region in $s$-wave SIS junctions at
mK-temperatures should have the zero flat region practically up to
the gap voltages $\pm 2 \Delta/e$, and exhibit a very sharp
conductance peak structure at $V= \pm 2 \Delta/e$. This is clearly
not in agreement with our results. Despite the fact that the
measurements were carried out at very low temperatures, the gap
structure is strongly smeared with a broad conductance peak, and
the shape of the tunneling spectra are not consistent with the
expected tunneling conductance for $s$-wave superconductors at
temperatures near 0 K. On the other hand, the quadratic in $V$
behavior of the subgap tunneling conductances observed in this
work is in agreement with the conductance of the SIS junction
calculated by Won and Maki \cite{Won94} with $d$-wave order
parameter, although the shape of our experimental curves at
voltages above the conductance peaks differs strongly from the
theoretical curves.

\section{Quasiparticle density of states}

In order to extract the quasiparticle DOS, $N(E)$, from the
tunneling conductances in Fig.\ref{fig3}(a-c), we deconvoluted the
expression for the normalized tunneling conductance \cite{Wolf85}

\begin{equation}\label{eq1}
\frac{dI/dV(V)_S}{dI/dV(V)_N}=\frac
d{deV}\int_0^{eV}N(E)N(E-eV)dE.
\end{equation}

\noindent This expression suggests that a tunneling matrix element
which weights a particular momentum direction is a constant and
the tunneling spectrum directly relates to the DOS in the CuO$_2$
plane. Although in some articles the tunneling spectrum was
considered to be suggestive of an anisotropic matrix element, in
Ref. [\onlinecite{Hoog03}] it was shown on the basis of a detailed
analysis of the tunneling and ARPES spectra that the tunneling
matrix element does not have a strong dependence on the wave
vector. Tunneling spectroscopy therefore equally probes states
over the entire Fermi surface.\cite{ Hoog03}

An iterative procedure for the deconvolution together with the
method used to obtain the normalized tunneling conductance is
described in Ref. [\onlinecite{Ved94}]. In Fig.~\ref{fig3}(d-f)
the data points (squares) represent the result of this
deconvolution for three different tunneling junctions from
Fig.\ref{fig3}(a-c), respectively. The solid lines show the
experimental conductance curves $(dl/dV)_{S}/(dI /dV)_{N}$
compared with the calculated curves (circles) using the $N(E)$
obtained from the deconvolution of Eq.(\ref{eq1}). The good
agreement with the data in Fig.~\ref{fig3}(d-f) confirms the
validity of the iterative procedure used to obtain $N(E)$. The
shape of the $N(E)$ showed in Figs.\ref{fig3}(d-f) is fully
compatible with a smeared BCS DOS for Bi2212.\cite{Ved94} The
obtained $N(E)$ over the entire subgap region can be fitted quite
well to curves which are quadratic in $V$ (dashed lines). A
quadratic rather than linear increase of $N(E)$ in the subgap
region seems to be strong evidence against $d$-wave symmetry.

Note that a quadratic or nonlinear in $V$ subgap tunneling
conductance in Bi2212 has been observed previously in STM
experiments, for example in Ref. [\onlinecite{Miyakawa99,Misra,
Renner}]. Franz and Millis \cite{Franz} have shown that a
reasonable fit to the data in Ref. [\onlinecite{Renner}] assuming
a DOS with a $d$-wave order parameter could be made provided the
tunneling matrix element is anisotropic. Assuming an isotropic
(constant) matrix element leads to sharper V-shaped gap structure,
inconsistent with the experimental data.\cite{Renner} As
previously noted, the tunneling matrix element is most likely
independent of the wave vector,\cite{ Hoog03} which implies that
the STM spectra give the quasiparticle DOS directly, and are
therefore inconsistent with a pure $d$-wave order parameter, in
agreement with our results. For this reason, we are forced to
conclude that it is not possible to explain our data in terms of
either pure $s$ pairing or pure $d_{x^2-y^2}$ pairing. Our results
lead to the conclusion that Bi2212, in all likelihood, has a $d+s$
mixed pair state, as has been repeatedly suggested in the
literature (see e.g. Ref.\onlinecite{Li99, Latyshev04}).

\section{Sub-gap structure}

\begin{figure}%figure 4
\includegraphics[width=0.9\linewidth,angle=0,clip]{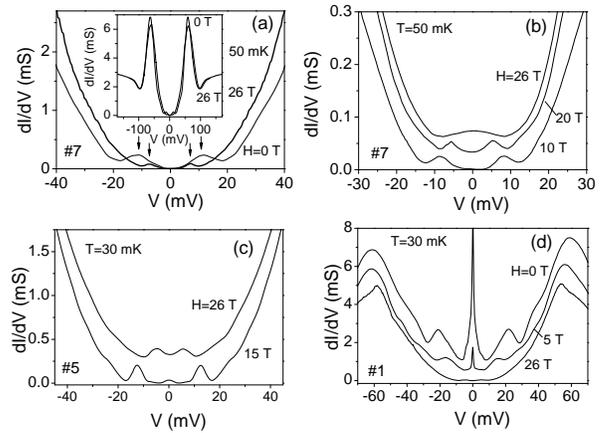}
\caption{\label{fig4} The effect of the magnetic field on the
gap-like structure (marked by arrows) in the tunneling conductance
$dI/dV$ vs. $V$ within the superconducting gap for different
break-junctions on three single crystals at $T=50$ mK and $T=30$
mK. (a,b) - two tunnel break-junctions with resistance nearly 1
k$\Omega$ fabricated with the same single crystal $\#7$. The inset
shows the full gap spectra at 0 and 26 T. (c,d) - the subgap
regions of the spectra on an enlarged scale for two single
crystals $\#5$ and $\#1$ (the resistances nearly 300 $\Omega$).
For clarity, the curves in (b-d) have been shifted vertically with
respect to the lower curves. }
\end{figure}

In addition, a large number of the tunnel break-junctions
exhibited $dI/dV(V)$ curves with gap-like features within the gap
at voltages $\sim 6-20$ mV. With increasing magnetic field this
structure broadens, diminishes in amplitude and shifts to lower
voltages. In Fig.\ref{fig4}, we show the subgap regions of the
differential conductances $dI/dV$ as a function of the bias
voltage $V$ in different magnetic fields at $T\simeq 50$ mK. (a,b)
- two tunnel break-junctions with resistance of nearly 1 k$\Omega$
fabricated with the same single crystal $\#7$ (the low-voltage
gap-like structure is indicated by arrows). The inset in (a) shows
the full gap spectra at 0 and 26 T in order to compare the
amplitude of the sub-gap features with the gap peak.
Fig.\ref{fig4}(c,d) shows the sub-gap regions of the spectra on an
enlarged scale for two single crystals $\#5$ and $\#1$ (with
resistances of around 300 $\Omega$). For clarity, the curves in
Fig.\ref{fig4}(b-d) have been shifted vertically with respect to
the lower curves.

It might be thought that the observed sub-gap structure are
low-energy features related to the zero-bias peak which is caused
by $d$-wave Andreev bound states \cite{Fogelstrom, Covington} or
the Josephson effect. However, in case of the higher-resistance
break-junction (Fig.\ref{fig4}(a)) the zero-bias peak is lacking
at $H= 0$ T. Nevertheless, both the sub-gap structure and its
field dependence are seen. In Fig.\ref{fig4}(b-d), the sub-gap
gap-like structure is also clearly visible despite the fact that
the zero-bias peak is suppressed completely by the magnetic field.
This data show that the zero-bias peak and the sub-gap gap-like
structure in our case have distinctly different origins.

Taking into account the magnetic field dependence of the gap-like
structure observed here, we identify this feature with a small
energy gap, the average magnitude of which is 2$\Delta^s
_{p-p}/e=13$ mV. Electronic structure calculations for Bi2212 have
long predicted the presence of BiO-derived bands and a splitting
of CuO$_2$ bands at the Fermi energy. If we assume that the
break-junction tunneling averages the DOS equally around the Fermi
surface,\cite{ Hoog03} then the observed tunneling conductance
must involve two separate order parameters derived from different
bands. While the sub-gap structure might be connected with a
proximity induced energy gap in the BiO layers,\cite{Yurgens} the
possible existence of a second small gap (2$\Delta \le 8-20$ meV)
in Bi2212 has been previously suggested by Shen \textit{et al.}
\cite{Shen93}, Mahan \cite{Mahan93} and Mallet \textit{et al.}
\cite{Mallet96} in order to explain the gap anisotropy observed in
ARPES and vacuum tunneling spectroscopy of Bi2212. It should be
pointed out that the weak sub-gap structure in Bi2212 has been
occasionally seen in the vicinity of the zero bias in STM spectra
at 4.2 K \cite{Liu, Renner} and break-junctions at 10
K.\cite{Miyakawa99} It is also possible that the sub-gap structure
detected by us at mK temperatures can manifest itself as a
``linear'' increase in the subgap region in the vicinity of the
zero bias as often observed in the STM spectra of Bi2212 measured
at $T=4.2$ K.

\section{Superconducting energy gap}

We now turn our attention to the magnitude of the superconducting
energy gap in the studied crystals. For the SIS junctions studied
here, the peak-to-peak separation of the two main maxima of the
$dI/dV(V)$ curves must correspond to 4$\Delta _{p-p}/e$. As can be
seen in Fig.\ref{fig3}(a-c), this value is different for the
different break-junctions formed using the same single crystal
($\#1$) and equals 96, 93, and 124 mV. The applied magnetic field
$H\parallel ab$-plane should decrease only slightly the
peak-to-peak separation in $dI/dV(V)$ curves due to the very high
parallel critical magnetic field in Bi2212. For sample $\#1$, we
find a surprisingly wide scatter of the size of the energy gap
2$\Delta _{p-p}/e$ ranging from 46 to 78 mV for the more than 50
different break-junctions investigated. Similarly, for two other
single crystals ($\#5$ and $\#7$), we obtained 2$\Delta_{p-p}/e =
50-65$ mV and 2$\Delta_{p-p}/e = 48-72$ mV, respectively.

\begin{figure} %figure 5
\includegraphics[width=0.7\linewidth,angle=0,clip]{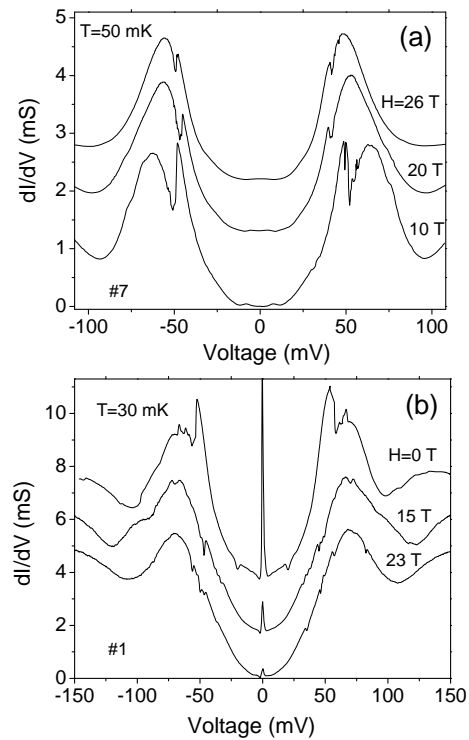}
\caption{\label{fig5} Differential conductances $dI/dV$ as a
function of the bias voltage $V$ for tunnel break-junctions
fabricated on single crystals $\#7$ and $\#1$ with double peaks in
different magnetic fields at $T\simeq 50$ mK. For clarity, the
curves have been shifted vertically with respect to the lower
curves}
\end{figure}

Miyakawa \textit{et al.}  \cite{ Miyakawa98, Miyakawa99} have
found a strong monotonic dependence the superconducting energy gap
of the Bi2212 on the doping concentration. However, we can
formally exclude macroscopic sample inhomogeneity, as the origin
of the observed variation, since 2$\Delta _{p-p}/e$= 46-78 meV
would imply \cite{Miyakawa98} a spread of hole concentration
$p=0.2-0.15$. Using the empirical relation for $T_c$ vs $p$, this
would imply values of $T_c$ ranging from 80 to 95 K. In addition,
the chemical composition of the crystal, measured at several
different points, is found to be identical (Fig.\ref{fig1}). The
crystal is of a very high quality judging from the sharp
superconducting transition width $\Delta T_c = 1.5$ K (in the
magnetization measurements), and the small rocking curve width
$\delta \theta \approx 0.1^{\circ }$. Following reference
\onlinecite{Kinoda03}, this strongly suggests that such a wide
range of the gap values originates from the presence of different
superconducting regions on the scale of a few coherence lengths.
The effects of a nanoscale chemical inhomogeneity on $T_c$  of
Bi2212 was recently studied in Ref. [\onlinecite{Eisaki}]. If the
sample has a nanoscale disorder with multiple energy gaps, the
convolution of several DOS gives a strongly smeared single
conductance peak (Fig.\ref{fig3}(a-c)), nevertheless, the DOS
after the deconvolution of the conductance can show an additional
maximum in the peak region (Fig.\ref{fig3}(f)). In individual
cases the break-junctions can show double peaks in the conductance
at voltages nearly 2$\Delta _{p-p}/e$ (Fig.\ref{fig5}), that is
indicative of two nanoscale superconducting regions in the region
of the junction.

\section{Conclusion}

In conclusion, we have studied the tunneling density of states
using high-quality break-junctions fabricated on Bi2212 single
crystals. Our experiments provide a crucial test for the presence
of excitations within the superconducting energy gap. The measured
tunneling conductances in the subgap region show a zero flat
region. We have never observed either a linear increase of the
tunneling conductances with voltage in a subgap region or a
nonzero zero-bias conductance at the temperatures 30-50 mK. The
deconvoluted DOS in the whole subgap region can be fitted quite
well to curves which are quadratic in $V$. A quadratic rather than
linear increase of $N(E)$ in the subgap region seems to be strong
evidence against pure $d_{x^2-y^2}$ symmetry. The major part of
the tunnel break-junctions exhibited $dI/dV(V)$ curves with the
second energy gap structure with an average magnitude
2$\Delta_{p-p}/e=13$ mV. We cannot explain our data with either
pure $s$ pairing or pure $d$ pairing. The wide range of the
superconducting gap values observed suggests that the results of
investigations of the doping dependence of the superconducting gap
in HTS by tunneling spectroscopy need to be treated with care.

\begin{acknowledgments}
This work has been partially supported by NATO grant PST.CLG.
979896.
\end{acknowledgments}

\bibliography{nnPRB}
\end{document}